# Photoluminescence properties of pyrolytic boron nitride.

**Luc MUSEUR**

*Laboratoire de Physique des Lasers – LPL, CNRS UMR 7538, Institut Galilée, Université Paris 13, 93430 Villetaneuse, France*

**Andrei KANAEV**

*Laboratoire d'Ingénierie des Matériaux et des Hautes Pressions – LIMHP, CNRS, Institut Galilée, Université Paris 13, 93430 Villetaneuse, France*

# Abstract.

We report on spectroscopic study of pyrolytic hBN (pBN) by means of time- and energy-resolved photoluminescence methods. A high purity of pBN samples (though low crystallinity) allows complementary information about excited states involved into the luminescence process. We affirm our recent conclusions about the impurity-related nature of most of the fluorescence bands in microcrystalline hBN. In addition, a broad band centred at 3.7 eV previously not considered because of its superposition with an intense structured impurity emission is attributed to the radiative recombination of deep DAPs.

# I. Introduction.

Hexagonal boron nitride (hBN) is a promising material for applications in optoelectronics. This stems from both its band-gap energy around 6 eV, one of largest of the III-nitride family, and from recent observations of far-UV (215 nm) laser action in single crystal under e-beam excitation[1]. In the same time despites of many experimental and theoretical efforts in the past, the electronic structure of hBN subjects to debates. Luminescent mechanisms in hBN appear ambiguous and need further investigations. In particular, a large diversity in shapes and positions of photoluminescence (PL) spectra published in literature is usually interpreted by sample preparation conditions and presence of impurities. Generally, impurity, defect and donor-acceptor (DAP) states may be formed within the hBN band-gap leading to luminescence in the visible or UV spectral regions. Therefore, it is of interest to analyze and compare luminescence properties of hBN samples with different structural defects and impurity contents.

Luminescence properties of high purity commercially available hBN microcrystal powders have been largely studied in cathodoluminescence[2-5] and photoluminescence[6-14] experiments. Due to the synthesis methods used, these powders are inherently contaminated by carbon, calcium and oxygen impurities. In contrast, pyrolytic hBN (pBN) which is obtained by chemical vapour deposition (CVD) method, is considered to be a more pure material [15]. From a structural point of view, pBN is a layered material composed of extremely small hexagonal crystallites (~few nm) highly oriented in the direction perpendicular to their basal planes, but randomly oriented within the layer. pBN can be considered as a basically 1D ordered material, whereas hBN powder microcrystals are arbitrary 3D ordered. Consequently, the density of stacking faults is higher in pBN than in hBN commercial powder.

In the present communication we report on a detailed experimental studied of defect related luminescence of pBN excited by synchrotron radiation (SR). Low-intensity SR enables observations of luminescence in conditions almost free of undesirable excitation effects. Conducted fluorescence experiments with energy- and time- resolution allowed better understanding of natures of the observed luminescence bands.

## II. Experiment.

Two kind of hexagonal boron nitride samples were analyzed. The first one was a pyrolytic BN (pBN) sample obtained by gas-phase interaction between $BCl_3$ and $NH_3$ at 2300 K and deposition on a Si substrate. This technique, in contrast to the traditional ceramic one, is free of carbon and calcium compounds and ensures, in principle, high purity samples [16]. The second sample was made from commercial hexagonal BN powders (Alfa 99.5%) compacted in a square pallet (8x8x1 $mm^3$) under a pressure of 0.6 GPa. The grit size of the hBN powder has been estimated by granulometry and transmission electron microscopy (JEM 100C JEOL). It ranged from 0.3 to 10 µm with an average particle size of 3.1 µm corresponding to the maximum in the mass distribution curve. Typical impurities contents, as provided by the Alfa Company, are listed in Table 1. The heating of hBN samples at 800 K under vacuum for a period of 12 hours did not provide significant changes in the PL spectra, which show low influence of the adsorbed surface species (as organic impurities and traces of water).

The luminescence properties of the samples were measured by means of the VUV SR excitation from the DORIS storage ring at HASYLAB (DESY, Hamburg). The facility of the SUPERLUMI station used for experiments is described in details in references [17, 18]. Briefly, samples were cooled down to 8K and irradiated by monochromatized SR ($\Delta\lambda = 3.3 \overset{o}{A}$) under high vacuum (~$10^{-9}$ mbar). Typical photons flux was $4.10^5$ photons/pulse (4 MHz) focused on a surface sample of 0.02 $cm^{-2}$. The measurements of luminescent spectra were carried out using a visible 0.275-m triple grating ARC monochromator equipped with a CCD detector. Spectra were recorded within a time gate $\Delta\tau$ delayed to the SR excitation pulse. Typically three time gates have been used simultaneously. In the rest of this article the time gates will be call in the following way: fast ($\Delta \tau_1 = 2-6\,ns$), medium ($\Delta \tau_2 = 10-30\,ns$) and slow ($\Delta \tau_3 = 60-200\,ns$). The recorded spectra were corrected for the primary monochromator reflectivity and SR current. The luminescence spectra were corrected for secondary monochromator reflectivity and CCD sensitivity.

X rays diffraction (XRD) spectra has been performed using a DMAX 2000 apparatus ($\lambda_{\alpha Cu}$=0.154056 nm). The samples were fixed vertically and the X-ray beam incidence angle was close to 90° with the normal to the sample surface.

## III. Results and discussion.

XRD spectra of both samples are shown on Figure 1. Well crystallized hBN powder exhibits a series of well resolved and sharp lines all assignable to hBN. In contrast, XRD lines observed with pBN sample are broader indicating smaller microcrystalline domains. The dimensions of microcrystalline domain can be calculated using the Scherrer's equation $L = K\lambda/(\Gamma \cos\theta)$, where K is a constant, $\lambda$ is the X-ray wavelength, $\theta$ is the diffraction angle and $\Gamma$ is the full width at the half maximum in radian. The planar domain size ($L_a$) is obtained from *hkl* lines with *K = 1.84*, whereas the domain extent perpendicular to the planes ($L_c$) is obtained from *00l* lines with *K = 0.9* [16, 19]. The obtained sizes are reported on Figure 1. Due to a poor resolution of the 100 and 101 lines, the $L_a$ dimension was not extracted for pBN.

A typical low temperature (9K) luminescence spectrum of pBN sample excited with 6.05 eV photons is displayed on Figure 2.a). Three luminescence bands are observed at energies of 5.3 eV, 3.75 eV and 3.1 eV. The very weak band at 5.3 eV has been previously reported in hBN microcrystal and assigned to quasi-donor-acceptor pair (qDAP) radiative transitions[6]. The most intense band at 3.75 eV (330 nm) is broad ($\Delta E \sim 0.6\,eV$) and has a rather symmetric shape. Conversely, the band at 3.1 eV (400 nm) exhibits an asymmetric profile with a long tail to the low energy side. These two last bands have been previously reported in pBN samples irradiated with charged particles [20, 21]. They were assigned to recombination of radiation induced free carriers with a quasi-continuum of impurities levels inside the band gap.

The photoluminescence excitation (PLE) spectrum of the 3.75 band is presented on Figure 2.b). This spectrum consists in two characteristic regions: (i) a broad shoulder for excitation energy ranging from $E_{exc}$ = 4.5 eV to 5.5 eV (ii) a peak centred at 5.8 eV. This peak is followed by a pronounced deep at 6.1 eV. We assign the peak at 5.8 eV to free exciton, since it lies between the free exciton luminescence observed in hBN single crystal at 5.76 eV [1, 22, 23] and the excitonic states predicted theoretically at 5.85 eV[24]. Unfortunately, because of an overlap between the low energy tail of the 3.75 eV band and the 3.1 eV band, it has not been possible to obtain a reliable excitation spectrum of this later.

The time-gated luminescence spectra of the 3.75 band are shown on the Figure 3. Three time gates defined in the part II have been used to resolve fast-, intermediate- and long- lived excited states. Our results indicate that this luminescence exhibits a red spectral shift when the delay with respect to a SR excitation pulse increases. In others words, the high-energy edge of this emission decays much faster than the low energy edge. In contrast, no shift or modification of the band shape has been observed for the 3.1 eV band.

These time gated luminescence spectra allow the assignment of the broad emission band around 3.75 eV to donor acceptor pair (DAP) radiative recombination. Indeed, the faster decay of the high energy edge of the band is characteristic of DAP transitions. The photon energy resulting from radiative recombination of a DAP at the distance $R_{AD}$ between donor and acceptor, is given by the following expression if the donor and the acceptor are not too close [25]:

$$h\upsilon_{DAP} = E_g - E_D - E_A + \frac{e^2}{4\pi\varepsilon_0\varepsilon_r R_{AD}} \qquad (1)$$

where $E_g$ is the band gap energy, $E_A$ and $E_D$ are the neutral acceptor and donor ionization energies with respect to VB and CB. Thus, the transition energy decreases with an increase of $R_{AD}$. Moreover, the probability tends decreasing with an increase of $R_{AD}$ because of a smaller spatial overlap between $A^0$ and $D^0$. As a result, the recombination at low photon energy (large $R_{AD}$) is slower than at high energy (small $R_{AD}$). This results in a red shift of the luminescence band when the time window is delayed with respect to the excitation pulse. Because of a similar behaviour, we assign the 3.75 eV luminescence band to the radiative recombination between neutral acceptor and donor states:

$$A^o + D^o \longrightarrow A^- + D^+ + h\upsilon_{DAP} \qquad (2)$$

The temperature dependence of the 3.75 eV luminescence band confirms this assignment. When the temperature increases, we observe a decrease of the band intensity and a modification of its shape without energy shift (Figure 2.a). The absence of energy shift is expected since changes of the band gap energy with temperature are supposed to be very small in hexagonal BN [26, 27]. Moreover, a careful comparison of the PL spectra shows that the thermal quenching is more efficient for the close DAPs, which contribute to the high-energy part of the band. This behaviour is characteristic of the DAP states, since closer neutrals (donor and acceptor) more shared the electron. This Coulomb interaction between the donor and the acceptor results in a lowering of their ionization energy. For close DAP the related decrease of the ionization energy can be sufficient for thermal release of trapped carriers and, therefore, the high-energy part of the PL band subjects to thermally quenched at lower temperatures. A similar effect has been previously reported in heavily doped GaN by Mg for the blue luminescence assigned to deep donor - shallow acceptor ($Mg_{Ga}$) transitions [28].

The PLE spectra presented on Figure 2.b), give insights about excitation mechanisms leading to the DAP luminescence. We assign the broad shoulder between 4.5 eV and 5.5 eV to the direct excitation of DAP luminescence by transition of an electron from an acceptor to conduction band or from the valence band to a donor:

$$\begin{aligned}A^- + D^+ + h\nu_{exc} &\to A^o + D^+ + e^- \to A^o + D^o \\ A^- + D^+ + h\nu_{exc} &\to A^- + D^o + h^+ \to A^o + D^o\end{aligned} \quad (3)$$

This conclusion is supported by recent photocurrent measurements in the single crystal, which have revealed that optical transitions in the range of 2.7-5.4 eV lead to a creation of free carriers [29]. Moreover, the shoulder can be successfully explained by the configuration coordinate model, which explains the coupling of electronic transitions to lattice vibrations. In framework of this model, a strong coupling to lattice results in a Gaussian absorption lineshape in agreement with the insert of Figure 2.b). Finally, the appearance of the excitonic peak at 5.8 eV shows that the energy transfer to the DAP states (luminescence at 3.75 eV) is also efficient:

$$exciton + A^- + D^+ \longrightarrow A^0 + D^0 \longrightarrow A^- + D^+ + h\nu \quad (4)$$

The pyrolytic boron nitride being a high purity material, the donor and acceptor involved in the 3.75 eV luminescent band are more likely related to intrinsic defects rather than to extrinsic impurities. Boron ($V_B$) and nitrogen ($V_N$) vacancies, which form respectively acceptor and donor levels around 1 eV from the band edge [30], may also account for the luminescence process. Experiments with pBN samples made with excess of boron or nitrogen could be useful to clarify this point.

The nature of the second asymmetric band observed at 3.1 eV in pBN samples is not elucidated by these experiments. In contrast to the 3.75 eV band, no shift or modification of this band has been detected in time-gated luminescence experiments (Figure 3) indicating a rather long lifetime (>200 ns). This seems exclude a possibility of free carrier recombination on impurity, which is expected to be faster. Interestingly, a similar broad band around 3.15 eV has been observed in cathodoluminescence spectra of boron nitride nanotubes [31] and on photoluminescence on hBN powder [32]. Its appearance in hBN powder has been related to the luminescence of a strongly self trapped exciton on the edges of hBN micro crystals where dangling bonds can form nanoarches (half nanotubes) between neighbour sheets of BN [32]. We note that pyrolitic BN is a solid material, inside which there is a strong contact between all nanocrystallites. Consequently the density of dangling bonds is expected to be lower than in hBN powder. It seems therefore difficult to assign the 3.15 eV to excitons trap on half nanotubes.

Luminescence spectra of usual hBN microcrystal powder are displayed on Figure 4 These spectra have been obtained in the same experimental conditions that pBN luminescence spectra and are discussed in details in recent papers [6, 13]. A comparison between luminescence properties of both kinds of samples brings information.

(i) The near band gap emissions are significantly less intense in pBN compare to hBN microcrystal powder. Particularly, the strong emission at 5.5 eV is not observed in pBN. Recently this emission in hBN has been assigned to bound exciton luminescence caused by disorder such as shearing of the lattice planes or stacking faults [2, 33]. Such defects are also present in pBN. However, because of the strong exciton localization within hexagonal planes [24], the observed quenching of the exciton luminescence is more likely related to non-radiative recombination on edges of the small nanocrystals ($L_c$ = 6 nm) composing pBN. Moreover, in a recent publication[6] we have assigned the 5.3 eV emission to quasi donor acceptor pair (q-DAP) states due to electrostatic band fluctuations induced by charged defects. A very low intensity of this band observed in pBN is consistent with the high purity level of this material.

(ii) A strong structured emission is observed in hBN under excitation energy $E_{exc} > 4.1\,eV$ (Figure 3.a). This emission has been assigned to impurity luminescence[7] and correlated with the presence of carbon and/or oxygen impurities in the sample[34]. Whatever the excitation energy, we have never observed this luminescence in pBN sample. This is a strong support to a carbon defect origin of luminescence as proposed in literature [34, 35].

(iii) In hBN microcrystal, a broad DAP emission around 3.9 eV is observed (figure 3.b). A large bandwidth ($\Delta E \sim 1 eV$) and nearly symmetrical shape of this emission has been explained by presence of deep acceptor complexes centers strongly coupled with the lattice[13]. Such localized centers, so called "A-centre", were supposed to be formed by an impurity atom, likely C or O, in association with an adjacent boron vacancy. Actually, in pBN sample, the DAP luminescence is more narrow ( ΔE ~ 0.6 eV ) and less extended to the high energy side. This conveys both a larger distance between acceptor and donor states, and a less efficient coupling to the lattice. Therefore we can conclude that acceptor and/or donor states involved in DAP luminescence process are different in both kind of sample. In particular, absence of carbon impurity in pBN could explains the smaller bandwidth of the DAP luminescence.

## IV. Conclusion.

The luminescence properties of pyrolytic boron nitride in the energy range 3 – 4 eV has been studied by means of the time- and energy-resolved photoluminescence spectroscopy methods. Thanks to the time gated luminescence we have assigned the broad luminescence observed around 3.75 eV to DAP radiative recombination. A comparison is made with hBN microcrystal powder luminescence. The strong structured UV luminescence, which is currently reported in hBN powder, is not been observed in the pBN samples, underlining importance of carbon impurities in the luminescence process. Moreover, bound excitons luminescence at 5.5 eV vanishes in pBN sample. This is ascribed to absence of shearing defect which usually act as radiative recombination centres in hBN microcrystal.

## Acknowledgments

This work has been supported by the IHP-Contract HPRI-CT-1999-00040 of the European Commission. The authors are grateful to G. Stryganyuk for assistance in conducting experiments at SUPERLUMI station and to V. Solozhenko for helpful discussions and the kindly providing of pyrolytic boron nitride samples.

**Tables :**

| Sample | C | O | Si | Ca | $B_2O_3$ | Fe | Al | Cl |
|--------|-------|------|-------|-------|----------|--------|--------|--------|
| hBN    | 0.005 | 0.02 | 0.004 | 0.011 | 0.14     | 73 ppm | 68 ppm | 29 ppm |

**Table 1 :** Impurities content of the commercial hBN powder in wt %.

**Figures :**

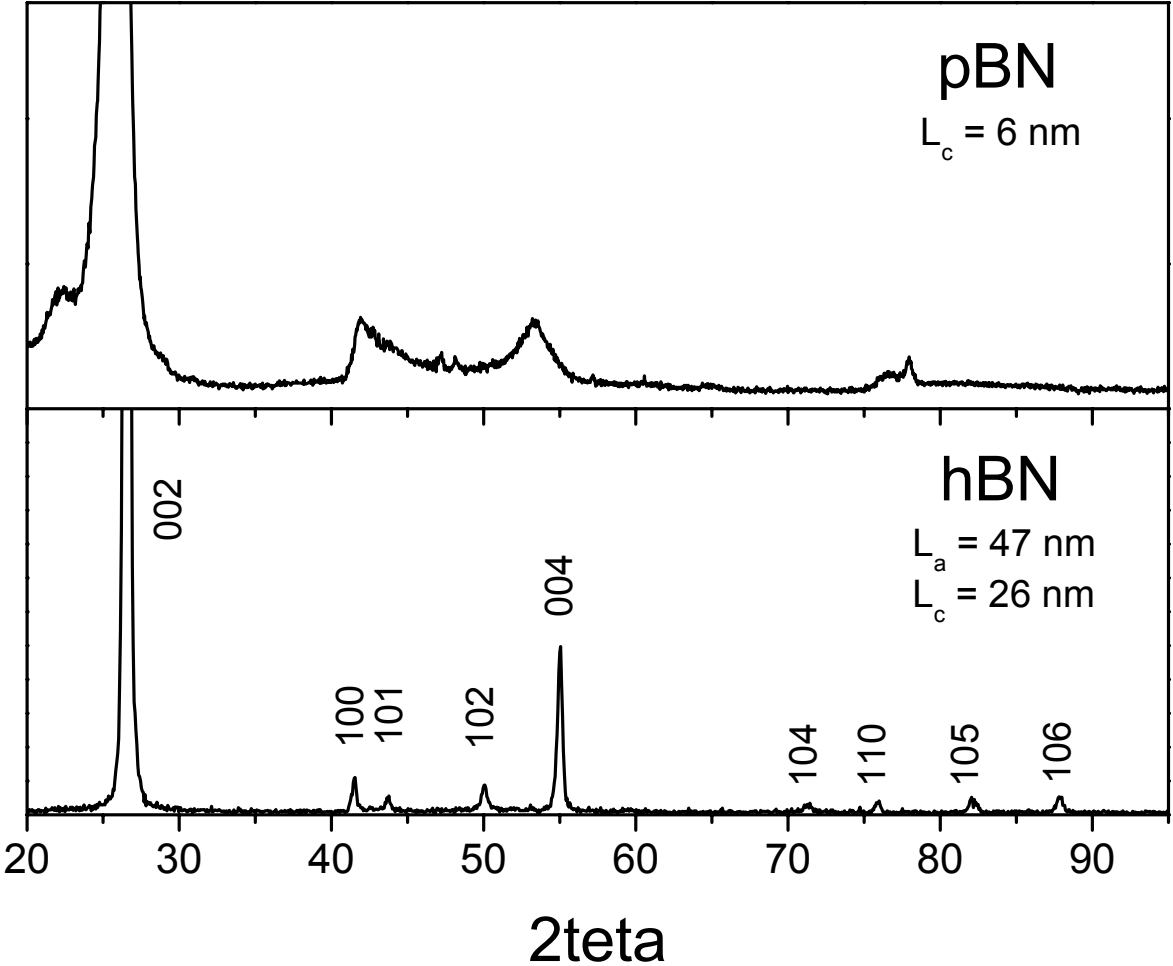

Figure 1. XRD spectra of pyrolytic BN (a) and commercial hBN powder (b)

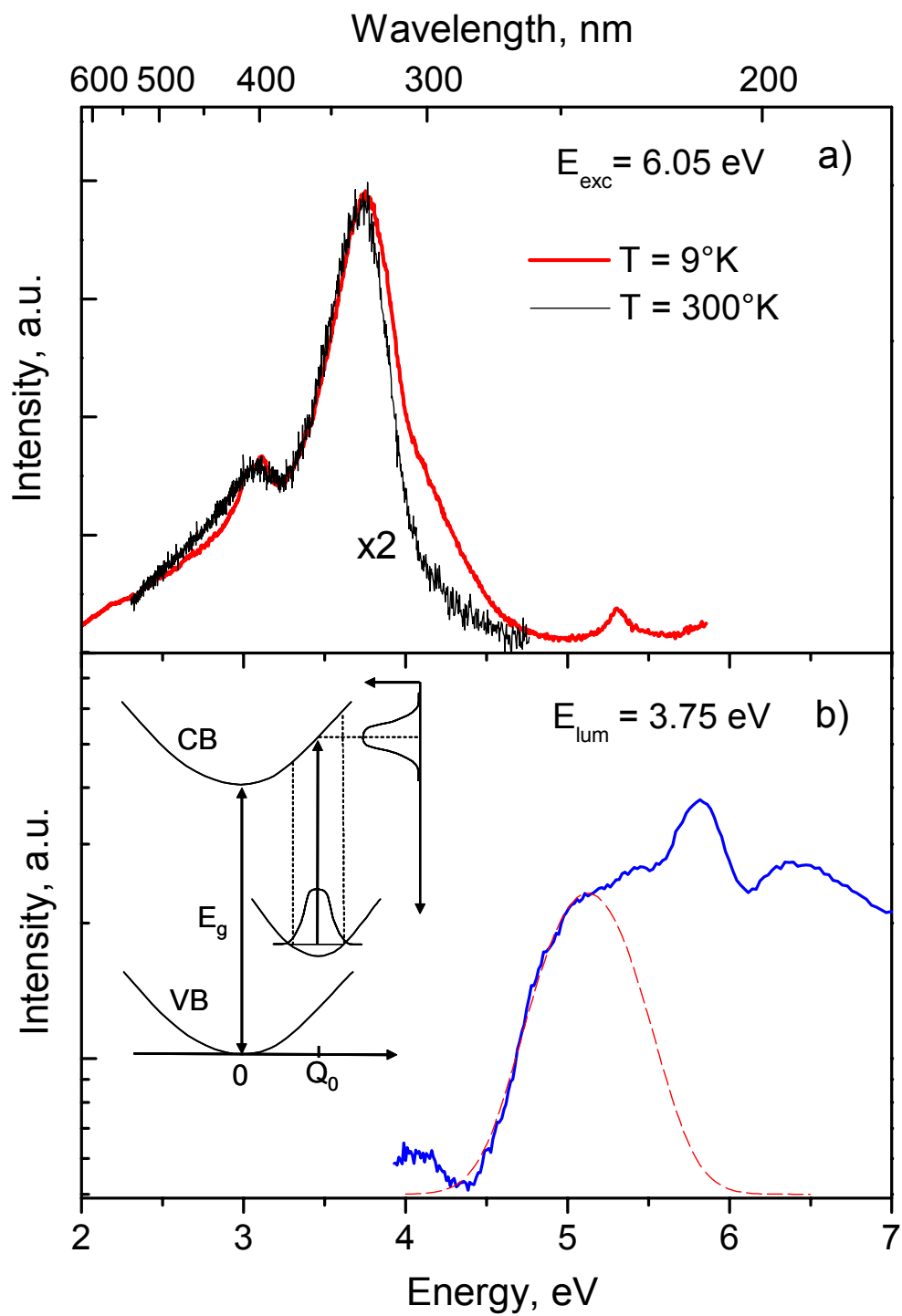

**Figure 2.** (a) Luminescence spectra of pBN sample (T = 9K and T = 300K) excited at 6.05 eV. To make easier the comparison, the two spectra have been arbitrarily normalized to their maxima (b) Photoluminescence excitation spectra at 3.75 eV. The dotted line shows a fit of the absorption edge by a Gaussian curve. The insert presents a schematic CC diagram showing the ionization of a deep acceptor.

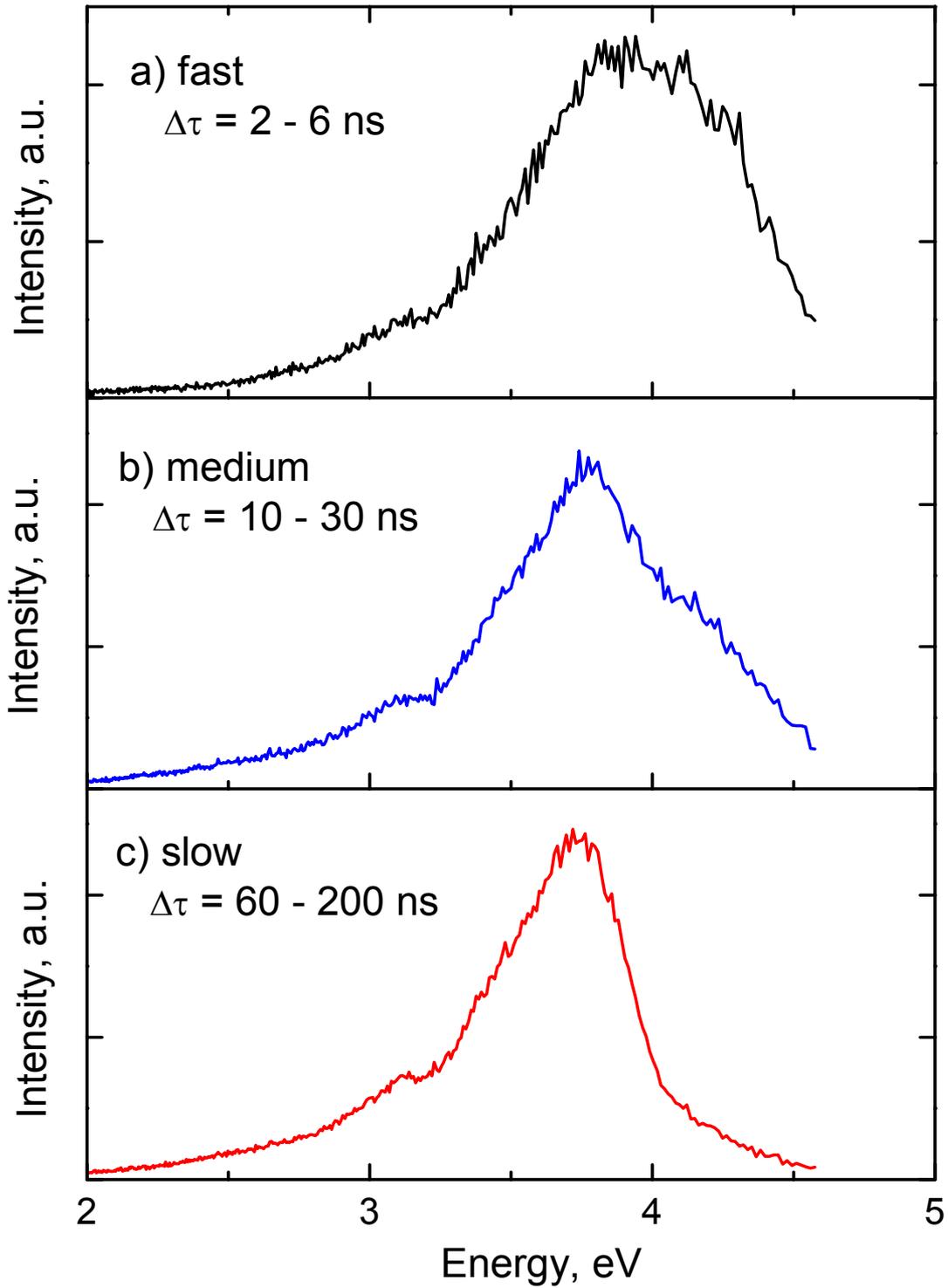

**Figure 3** Luminescence spectra of the pBN sample (T = 9K) registered with three various time windows delayed from the excitation pulse: fast ($\Delta\tau_1 = 2-6$ ns), medium ($\Delta\tau_2 = 10-30$ ns) and slow ($\Delta\tau_3 = 60-200$ ns).

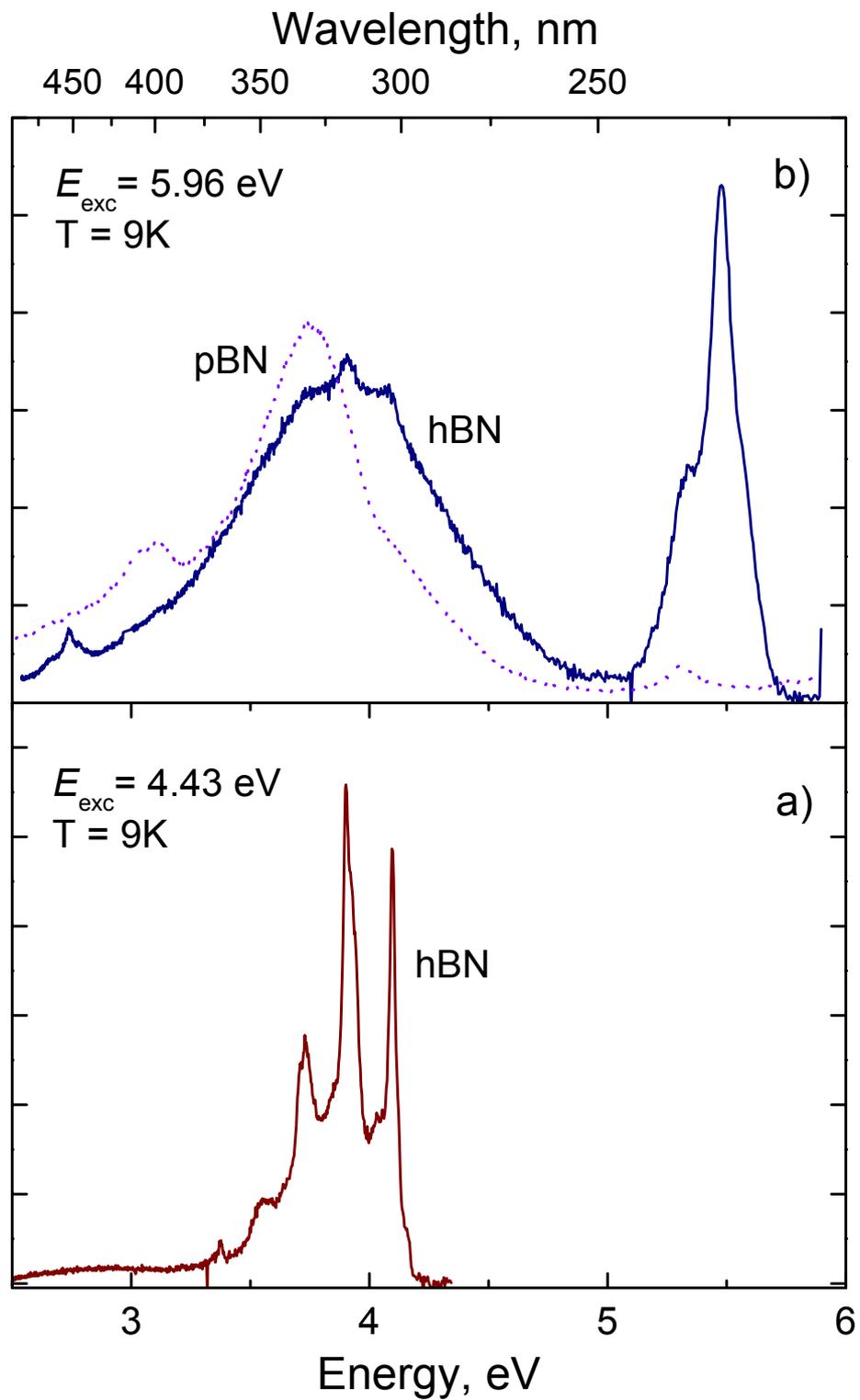

**Figure 4**: Photoluminescence spectra of hBN powder at excitation energy (a) $E_{exc} = 4.43\,eV$ and (b) $E_{exc} = 5.96\,eV$. For the sake of comparison luminescence spectra of pBN has been displayed on (b)